# Coherent and Noncoherent Photonic Communications in Biological Systems

## Mayburov.S.N.,

*Lebedev Inst. of Physics, Moscow, Russia*
*E-mail: mayburov@sci.lebedev.ru*

**Abstract**

The possible mechanisms of communications between distant bio-systems by means of optical and UV photons are studied. It is argued that their main production mechanism is owed to the biochemical reactions, occurring during the cell division.. In the proposed model the bio-systems perform such communications, radiating the photons in form of short periodic bursts, which were observed experimentally for fish and frog eggs[1]. For experimentally measured photon rates the communication algorithm is supposedly similar to the exchange of binary encoded data in computer net via optical channels

### 1.Introduction

Currently, the term 'biophoton' is attributed to the optical and UV photons emitted by bio-systems in the processes which are different from standard biochemi-luminescence. The significant bio-photon production (BP) in optical and close UV range is established for a huge variety of bio-systems[1,2]. Its energy spectrum is nearly constant within optical and close UV frequency range, so it is essentially different from the spectra expected for the system with the temperature about $300^o$ K, which in this range should fall on 15 orders of magnitude[2,3]. The related phenomena is the delayed luminescence (DL) which is the secondary photon emission, stipulated by the irradiation of the biological sample by the short pulse of visible light. The observed DL time dependence is strikingly different from the exponential one, expected for the independent emission of photons. BP rate demonstrates the nonlinear dependence on some bio-systems parameters, such as the density of living cells in the culture, etc[3]. The experiments evidence that BP rate is proportional to the rate of cell mitosis in particular bio-system[1,3]. The novel feature of bio-photon radiation is that, despite its low rate, about 10 photons/cm$^2$sec, it effectively performs the communications between distant bio-systems. In particular, being radiated by the developing bio-system, it can rise the rate of cell mitosis in another bio-system of the same or similar specie up to 50%, this phenomenon called mitogenetic effect (ME)[1,2]. The communications of some other types were reported also, for example, for the bio-systems in the state of abrupt stress[1] or slow destruction[10].

Until now, BP and DL properties, as well as ME, can't be described within the standard premises of cellular biology[2,3]. To explain them, Popp proposed that the electromagnetic (e-m) field of bio-system is in coherent quantum state, analogously to laser[3]. Such field can produce the nonlinear patterns in some situations, but only for a very short time. However, the observed range of DL lifetimes, up to several minutes, are hardly compatible with the quantum coherence (QC), because such field inevitably suffers the fast loss of coherence from its interaction with the molecules of bio-system and its environment. For typical bio-systems the field coherence should be destroyed during several picoseconds, the recent experiments confirm this predictions for optical excitations in bacteria[4]. This and other objections put the serious doubts on the hypothesis of long-living coherent e-m field in bio-systems. Note however, that this objections don't exclude the presence of short-time (about 1psec) or spacious coherence of biophoton field, some experiments demonstrated such effects [3,11].

Alternatively, in this paper we propose that BP nonlinearities and ME can be explained mainly by the well-known properties of living organisms, in particular, their complex dynamics which can result in the nonlinear reactions on the external perturbations and signals. Such back-reaction can induce the emission of response radiation, which in its turn can influence the behaviour of initial radiation source and result in the complicated nonlinear correlations between the source and receiver. The exchange of information between the bio-systems or between the distant parts of the same bio-system is analyzed in the framework of information theory. Basing on it, the simple phenomenological model of information exchange between the bio-systems by means of noncoherent photons was proposed[13]. It is shown that the many features of its mechanism are similar to the communications between computers by the binary encoded messages. The quantum-optical calculations of bio-system excitation explains also the influence of external nonbiological irradiation of bio-systems on their BP rate and ME[10]. This model, in principle, can incorporate also the effects of short-time coherence of biophoton field, yet in this paper we concentrate on more simple noncoherent case.

## 2. Information Exchange inside Bio-Systems

Before considering the photon exchange between the bio-systems, it's worth to consider how the similar communications can be realized between the distant parts of the same bio-system. The optical and UV excitations in the dense media can exist and spread as the quasi-particles called excitons $\varepsilon$ with mass $m$ which can spread freely through the whole media volume[5]. They are strongly coupled with e-m field, so they can be produced during the photons absorption by the media with high efficiency, the inverse process results in the photons emission from the system volume. It's established experimentally now that the excitons play the important role in the energy transfer inside the bio-systems, in particular, during the photosynthesis in plants and bacteria[6]. BP mechanism related to the nonlinear excitons spreading along the protein molecules was considered in[7]. In our model the excitations of biological media as the whole play the main role in biophotons generation and absorption[13]. We don't consider here any model Hamiltonian of excitons, which is planned for the future, however, for such long distances it's inevitably should have the solitonic properties[6]. DL experiments show the presence of IR and optical excitations in biological media, which rate is many orders larger than for the thermal equilibrium. Hence it can't be excluded that the energy losses of exciton due to its dissipation and rescattering can be compensated from the interactions with such excitations which transfer their energy to the exciton. Thereon, we suppose that the excitons spread freely over all the volume of bio-system. Under this assumptions, the exciton signalling between two parts $A_{1,2}$ of the same bio-system and photon signalling between two bio-systems A,B can be quite similar, so the mechanism proposed below for exciton signalling is applicable also to photonic one with minor changes.

As was noticed above, BP rate is quite low, about 10 photons/cm$^2$sec from the surface of large, dense bio-system. If such e-m field isn't coherent, then it described as the stochastic ensemble of photons. Then at its best the absorption of single photon or narrow bunch of photons can be detected by the bio-system as the single independent 'click' or one bit of information, analogously to standard photodetector devices. This is the photocounting regime of e-m field detection well-known in quantum optics[8]. We assume that the same approach is applicable also for the excitons radiated and absorbed in the different regions $A_i$ of the same bio-system. For example, the exciton can be generated in $A_1$ and to be detected as the single 'click' in $A_2$ at some distance from $A_1$. For the typical ME conditions in each cell cycle one absorbed bio-photon on the average corresponds to $10^{1 \div 2}$ cells in which it supposedly induces or accelerates mitosis[1]. Of the similar order should be this ratio for the excitons, hence the absorption of exciton by the individual cell seems to be too crude approximation of the real ME mechanism. Probably, the absorption mechanism is more similar to the simultaneous collective interaction of exciton with many nearby cells of the same region $A_i$, i.e. the cell cluster $r_j$ which can be regarded as the minimal bio-system region interacting with the excitons. For the simplicity it assumed in our model that any bio-system can be presented as the finite set of nonintersecting cell clusters. Each absorbed exciton supposedly is identified by the cell cluster $r$ as the single short-time 'click'. Hence after exciton absorption each cell of cluster supposedly receives the signal which can initiate or accelerate its mitosis. We shall suppose that on the average, the mitosis rate $R_M$ in a given cell cluster $r$ grows proportionally to the number of excitons $N_e$ absorbed in it. The size and other properties of cell cluster probably depends on bio-system sort. For fish eggs, which mainly are considered here, its size can be of the order of .1 mm, this is close to loach fish egg diameter and in some situations such egg probably interacts with impact biophoton as the single collective object[1].

It is well known that the cell mitosis is the complicated, multi-step process, during which the huge amount of genetic information is transferred in 2-bit nucleotide encoding to the chromosomes of new cell with minimal errors rate. It was shown that the functioning of cell nucleus in many other aspects is similar to standard microprocessor[6]. Hence the recognition of the external signals by the cell, which initiate or control its mitosis, probably can be quite similar to computer algorithms. Basing on this assumptions, the simple scheme of communications between the cell clusters or bio-system regions can be proposed. This scheme can be deduced from the first principles of information theory, but its the formal derivation is rather tedious, yet it can be obtained from the simple common sense reasoning. It's quite similar to the standard exchange of binary encoded messages between two computers $C_1$ and $C_2$ performed via noisy communication channels. Plainly, for low exciton or photon radiation rates, typical for bio-systems, the primary important problem is to suppress the background induced by all possible sources, like the violations of cell metabolism, the natural radioactivity, etc. In this case, as the approximate criteria characterizing the efficiency of information exchange, the signal to noise ratio $K_O$ can be used, i.e. the ratio of registered 'clicks' induced by bio-system signals and the background. It's natural to expect that the evolution of living species made the information exchange by means of excitons or photon radiation/absorption practically optimal for given bio-system with the limited average radiation rate. Since the time distribution of background radiation is constant, under this condition the optimal way for bio-systems to achieve high $K_O$ level is to make the main bulk of the bio-system radiation to be concentrated inside the short time periods, i.e. the bio-system radiation should form the periodic or quasi-periodic bursts which encode the meaningful information transferred to other bio-system. Its recognition and decoding supposedly can be performed by the methods similar to the standard ones used for noisy computer channels[13].

To illustrate them, we consider the selection and recognition of simplest bio-system signals. Such signal constituted by the long sequence of narrow bursts with average amplitude $I_g$ and period $T$ starting at some $t_0$. $I_g$ and $T$ distributions are taken to be Gaussian with relative dispersions about 20 %. The experiments evidence that the similar signals initiate and regulate the self-similar bio-system development, i. e. the multiple nonstructurized cell division, which occurs during yeast growth and fish egg cleavage[1]. In the presence of significant background which amplitudes are comparable with the useful signal the registration of single 'click' with amplitude higher than some fixed threshold would give $K_O$ value which is too small for effective noise suppression. Under this conditions the simplest recognition template (algorithm) can be $RT_1$ which demands that $N$ or more bursts with amplitude $I$ larger than some threshold $I_b$ are registered during fixed time interval $T_{max} \approx NT$; it can be called the repeatable or multiple bursts detection. More fine template $RT_2$ demands in addition that such bursts should be periodic up to $T \pm \sigma_t$ where $\sigma_t \approx .2T$. Such 'time encoding' of exciton bursts, in principle, permits to rise the final signal/noise ratio $K_f$ by the additional large factor $(K_O)^N$ without loss of detection efficiency. We found that despite that, normally, the optical noise is large, such templates permit to select the periodic bio-system signals effectively choosing suitable values of $N$ and other template parameters. Note that for the realization of $RT_2$ algorithm, the bio-system or the cell cluster, beside the registration of exciton as the single 'click', should also record the time of this event with the accuracy about 1 sec. However, the existence of 'biological clocks' in the living organisms, which work in quite different time ranges, is well established now. There are multiple evidences that for many bio-systems, for example yeast, such biio-communications are beneficial for their germination[2]. In particular, such long-distance communications inside the bio-system, can, probably, explain also the morphogenesis of developing bio-systems[11].

Let us suppose that in addition to this 'main' radiation cycle with period $T$, the bio-system radiation contains also other longer cycles with periods $T_L$. They wouldn't interfere with each other only if their periods are related to the shortest period $T$ as $T_L=LT$, where $L$ are arbitrary natural numbers, because only in this case, the bursts related to the different cycle $L$ can occur at one of the moments $t_j = t_0+jT$ of main cycle, but divided by interval $T_L$. For any other period relations the bursts related to one cycle would constitute the noise for the identification of other cycles. Such additional burst cycles, to which corresponds the extra harmonics in Fourier time spectra of radiation, can contain, in principle, the additional information which regulates the fine features of bio-system development, like the cells specialization in particular bio-system region, etc.. The recent results for BP in fish eggs and fibroblast cell cultures show the similar time structure for biophoton production[1]. In the measurements of optical radiation from fibroblast cells the short periodic photon bursts were revealed, they correspond mainly to the periods $T \approx 10, 40, 100$ and $400$ sec. For loach fish eggs the most pronounced harmonics of time spectra are 7, 70 and 280 sec. For both species the relative intensity of harmonics can vary depending on the development stages and external interventions into their development, like the compression or injection of drugs, stimulating the development. However, the presence of periodic bursts, which periods are related by the natural numbers is the invariable feature of all obtained spectra. Preliminary calculations, based on the experimental levels of noise and signals for fish eggs, show that demanding the efficiency of bio-system signal selection *90 %* , for the signal recognition by algorithm $RT_1$ one obtains $K_O=45$ for N=6. For algorithm $RT_2$ and N=4 it gives $K_O=3.2*10^2$; hence both this algorithms result in effective noise suppression, which practically excludes the development initiation by the experimentally observed level of photonic noise. Note that both algorithms work out the yes/no decision on the presence/absence of useful signal in incoming radiation during the period $D_T$ about 1 min, which corresponds closely to the admissible spread for the beginning of cleavage stage for individual fish eggs of the same eggs colony.

If the cell mitosis is stipulated by the synchronized regular bursts of excitons, than it's sensible to expect that the bio-system should contain the auxiliary mechanism which permit to vary slightly the time of exciton emission by the individual cells to conserve their synchronization. Really, the external perturbations and internal bio-system fluctuations would tend always to destroy this synchronization in the long run. Hence the bio-system should perform the program which permanently monitor the burst moments in its distant parts and restore their synchronization.. From this reasoning we shall assume also that the bio-system as the whole, as well as its parts or cell clusters, can also vary $T$ value for some small $\Delta T$ with each new burst, it is enough that it can be of the order $10^{-2}T$. As was supposed above, the cell clusters can 'measure' the exciton rate $R_e(t)$ with significant time precision. Because of this sensitivity and assumed tendency to the synchronization with irradiation peaks, under the external irradiation the burst moments $t_j$ of organism each time supposedly can be shifted in the direction of larger radiation intensity. The simplest example of such tuning program is the linear algorithm. In its framework, if this irradiation is periodical with the same period $T$ and its maxima lays at:

$$t_j^M = t_j + aT \qquad (1)$$

where *a<.5,* then $\Delta T>0$ and the moment of cell burst would shift in the direction of the irradiation maximum so that:

$$t_{j+1} = t_j + T + \Delta T \qquad (2)$$

until the external and internal radiation maximum would coincide in time[13].

### 3. Photon Communications between Distant Bio-systems

Now let us consider in this model approach the exchange of information by means of photon bursts between two distant bio-systems A, B of the same sort and age. Naturally, for both A, B the main period of exciton bursts has the same $T$ value, but by itself without temporary electromagnetic contact their burst moments will be independent: $t_i^B \neq t_j^A$ for arbitrary *i,j*; here we shall neglect the possible effects of higher time harmonics. Due to this time difference $t_d$ between A,B bursts, the photons radiated from A surface would not induce the mitosis in B and vice versa. Let's suppose that from some moment A, B are located nearby and can irradiate effectively each other. Then, under our assumptions those B bursts, which is the external irradiation for A would shift its moments $t_j^A$ of bursts for $\Delta T$ in each cycle in the direction of B bursts and vice versa. In the linear synchronization algorithm, described above, these variations of A, B periods will continue until both A, B bursts will be completely synchronized, which will take approximately:

$$t' \approx \frac{t_d T}{\Delta T} + T \qquad (3)$$

After that the normal $T$ value will be gradually restored for both A and B without violation of this synchronization. After the synchronization of A, B bursts, the photons radiated by A can be absorbed by B cells and produce the excitons in B during its burst period only. Those excitons can accelerate the cell mitosis in B and vice versa, i.e. A, B constitute the joint mitotic system. The considered A,B communications and their resulting transformation can be regarded as the simple example of back-reaction correlations. The experiments with the swarms of *Dinoflagellates* confirmed such effect and permitted also to study how such synchronization of flickering moments occur[3], the transition time for its establishment is less than 30 sec. Similar results are obtained for the bursts synchronization between two fibroblast cell cultures and two sets of fish eggs[1].

Now we turn to the study of communications between arbitrary bio-systems A,B, which structures and algorithms can be, in general, quite different. Plainly, this is quite complicated problem and here we only sketch the possible approaches to its solution, prompted by the analysis of experimental results. First, practically for all studied bio-systems the described burst structure of BP time spectra was found, yet their parameters depend on the age and other characteristics of bio-system. Altogether, the bio-systems are quite sensitive to such radiation and, in most cases, being irradiated for the period more than 10 min, fulfil strictly and obediently the commands transferred by it, despite that their realization can result in the damage and violation of their normal development or even destruction of bio-system. The illustrative example of it is the observation of programmed death – apoptosis of healthy epithelium cells induced by the radiation from the identical cell culture during its gradual destruction by the injection of $H_2O_2$ solution[10]. It seems possible that the time spectra of this radiation can be similar to the spectra measured for bio-systems in other stress conditions[1].

Now let us discuss the possible information content of signals which are transferred by this radiation. Here we shall consider only the case when A,B are of the same specie, but can be of different age $t_A, t_B$, which is classified usually by the development stage $n_A, n_B$. The experiments with fish and frog eggs evidence that the fine structure of A time spectra, in particular, the intensity of higher harmonics and related parameters of its Fourier spectra, depends strongly on $t_A$. Our main hypothesis is that A time spectra at development stage $n_a$ transfers the instruction which for the bio-system B dictates to perform the development program corresponding to the transfer of $n_A$ to the next stage of $n_A+1$, in particular, for the corresponding local rate of cell mitosis and cells specialization. Plainly, if $n_B$ differs much from $n_A$, the realization of such program can be quite unfavourable for B development in general.

Let us suppose first that $t_B<t_A$, their difference $t_{A,B}$ is relatively small and starting from some moment $\tau=0$ they are in close optical contact. Under the model assumptions, described above, at necessary B signal intensity A supposedly will develop as if it was at development stage $n_B$, but because $n_A, n_B$ difference is small and development processes at this stages possess the significant similarity, it means that, mainly, A development will just slow down. In the same vein, under A irradiation B development will be more similar to $n_A$ stage, hence B will slightly gain its speed of development. Consequently, when some time $\tau$ would pass, A,B development will be synchronized, as the result of such back-reaction communications. The results for the optical contacts during 20-40 min. between the samples of fish eggs A,B of slightly different age demonstrate the significant synchronization of their development[1]. Note that for the eggs colony produced by the fish at once it will be favourable that all eggs would develop with the same speed. However, the small variations of temperature and water flow over colony volume and other external and internal factors tend to destroy it. It seems that the biophoton signalling between distant eggs intend to restore their simultaneous development.

Now consider the situation when $t_A, t_B$, difference is large and hence A.B development processes differ much. In this case, any transition of A development to $n_B$ stage initiated by the signal of B radiation can deform A final structure seriously, the same is true for B under A irradiation. Nevertheless such deformed development, in principle, can proceed, at least for some time. During its realization, one can expect that A radiation time spectra to large extent will simulate the radiation at $n_B$ stage also (vice versa for time spectra of B radiation). The experimental studies demonstrate that the optical contacts during $20 \div 40$ min. of fish eggs of two significantly different ages result in the serious deformation of development in both samples, for fish eggs at early stages the development can simply stop. Correspondingly, in all this cases, the radiation rates form A,B are also reduced. However, the described effect of 'time spectra exchange' for two eggs samples A,B, which were in contact about 30 min,. was observed[1] in agreement with our assumptions.

### 4. Biophoton Dumping by Nonbiological Radiation

The multiple experiments evidence that by itself, the noncoherent optical radiation from nonbiological source, for example, the daylight, in general, has no influence on isolated bio-system[2,9] ( not accounting photosynthesis and other specialized processes). Yet in the presence of other bio-system the situation becomes more intricate and here we shall discuss its possible premises. DL experiments show that the typical bio-system possesses the large density $W(E)$ of metastable optical and UV levels, it seems that $W$ changes with energy $E$ rather insignificantly. For the bio-system in the darkness, considered until now, such levels are excited by some chemical reactions which on the average occur during bio-system development with nearly constant rate. The consequent photon emission in the form of periodic bursts with the average intensity $C$ is controlled by some collective mechanism of unknown origin. However, it is sensible to expect that, taking into account the large excitations lifetimes, its functioning is independent of the method of levels excitation. Therefore, if the same bio-system is exposed to the constant light flow $J$ with rather constant spectral distribution in optical range, for example the daylight, then the resulting amount of excited bio-system levels will enlarge in comparison with bio-system in the darkness[8]. Hence under such 'perturbed' conditions the bio-system should emit the photon bursts which amplitude $I$ will be, on the average, proportional to $C+kJ$ where $k$ is the fixed coefficient. It's reasonable to assume that for arbitrary bio-system the regarded threshold $I_b$, which define the burst registration and recognition, can be variable and adopts the level, optimal for signals recognition by the bio-system at given $J$. For the photocounting regime used in our model, the background should have Poisson distribution[8] and its rate, i.e. the number of background bursts of the height $I>I_b$ per time unit would be proportional to $\sqrt{J}$ for any fixed threshold $I_b$. Hence if the biophoton signals are selected by the algorithm similar $RT_{1,2}$, then such irradiation can be favourable for ME since signal/noise ratio $K_O$ would grow as $k\sqrt{J}$ as well, which would result in the background suppression with simultaneous gain of bursts amplitude. As $J$ would grow further, the levels saturation can become important, resulting in $K_O$ reduction and, probably, ME rate as well. It was found that at early stage of its development the fish eggs practically don't produce the photons. However, the constant background irradiation of low intensity is effectively absorbed by them, and main bulk of it is seemingly reradiated in form of periodical bursts[1]. In the same vein, the experiments show that ME for most of studied bio-systems is more pronounced when they are irradiated by the external visible light, than in case of complete darkness[10,12]. The optimal $J$ corresponds to about $10^{-1}$ of full day light intensity, for higher $J$ the rate of ME gradually declines[12].

Some experiments[10] evidence that ME is induced mainly by UV photons in the range from 200 to 300nm. Given it's correct, then in addition to the regarded 'direct' level dumping, the mechanism similar to the photons up-conversion can play the important role in ME. For example, if both bio-systems are irradiated by monochromatic flow $J$ with energy $E_1$ in optical range, it will excite the corresponding level $E_1$ of detector bio-system, and if one of them would absorb the biophoton $E_2$ emitted by the inductor bio-system, then the resulting excitation with energy $E_1+E_2$ will belong to UV range and initiate the mitosis with the larger efficiency. The similar effect will be observed, naturally, for the irradiation of detector bio-system by the continuous energy spectra with density $P_1(E_1)$ and biophotons from the inductor with spectral density $P_2(E_2)$. For large $J$ the resulting excitation spectra of detector bio-system is:

$$P_F(E) = \iint k_A(E_1) P_1(E_1) P_2(E_2) \delta(E - E_1 - E_2) dE_1 dE_2 \qquad (4)$$

here the integrals are taken over optical range, $k_A$ is absorption coefficient of bio-system B which is taken to be constant in optical range. DL experiments show that $P_2(E)$ is practically constant in optical range[2,3]. For $P_1$ corresponding to day-light spectra, one obtains the triangle $E$ spectra $P_F$ with the pick at

$$E_{MAX} \approx \overline{E}_1 + \overline{E}_2,$$

the resulting excitation energy $E$ would extend effectively to UV range. Note that at high intensity of radiation in UV range, the cells start to degrade and their mitotic mechanism can be spoiled as well.

## 5. Conclusions

In this paper it was shown that the exciton exchange supposedly constitutes the effective system of signalling and regulation of the bio-system development. It seems that such signalling to the large extent regulates the homogeneity of bio-system growth, preventing from the large fluctuations of its global form, i.e. defines its morphogenesis. In our approach, each cell cluster is the analogue of computer, which reaction is defined by the signals dispatched by the absorbed excitons[13]. Basing on it, the simple scheme of information exchange inside the bio-system was proposed, from that the similar scheme of photons communications between the distant bio-systems was derived. It turns out to be analogous to the standard procedure of information exchange between the distant computers by means of photonic signals transferred by the optical fibres. It's important to notice that the obtained scheme of photon communications is practically independent of particular BP mechanism. The calculations of BP time spectra in our model are in a reasonable agreement with the experimental results for BP in fish eggs[1]. Exploiting the rules of quantum optics, this model explains qualitatively the influence of external nonbiological irradiation on BP rate and ME . Note that this model doesn't demand that e-m field of bio-systems will be coherent during the long time periods, it permit to obtain ME and other biophoton effects assuming the produced e-m field to be noncoherent.

Yet it's worth to consider also the possibility that this e-m field of bio-system can possess the spacious short-time coherence within the observed photon bursts, similarly to the field coherence within the laser pulse. At least one experiment evidences directly that such coherence really takes place[11]. In its experimental set-up the transparent quartz plate was installed between the inductor and detector bio-systems. In the first run, the plate parallel surfaces were smooth and polished, so that it doesn't deform the phase relations between the different pieces of wave front of incoming photons. In another run, the plate has the random deflections from the surfaces parallelity, which violated such phase relations, and so violated the impact wave coherence. It was found that in comparison with the control sample of isolated bio-system, the radiation passed through the random surface results in the gain of mitosis rate of 20%, yet for smooth surface it reach the rate of 45%. In our framework, it's reasonable to suppose that e-m field with such 'transversal' coherence more effectively excite the collective excitations in the cells cluster, than the e-m field in which such phase relations are violated. If this hypothesis is correct, then such influence of coherence will not change the principal scheme of communications proposed here, rather, it would enlarge its efficiency.

The similar cell's control and regulation features are well studied for inter- and extra-cellular biochemical reactions[6]. Concerning the extracellular chemical signalling in the tissues, its efficiency and precision is principally restricted by the molecular diffusion effects inside the bio-system media. Note also that the exciton signalling inside organism can be much faster, than the corresponding chemical one. Hence it can be efficient in case of stress, or the abrupt change of external conditions. Experimental results show that under the different stress conditions the photon rates from bio-system can rise in short time significantly, probably, as the consequence of intensive internal signalling[1,2].